\documentclass[journal=ancac3,manuscript=article]{achemso}

\usepackage{chemformula} 
\usepackage[T1]{fontenc} 

\usepackage{float}
\usepackage{caption}
\usepackage{subcaption}
\usepackage{ulem} 

\newcommand{\pta}{\textcolor{black}}
\newcommand{\rya}{\textcolor{black}}
\newcommand{\ry}{\textcolor{black}}

\usepackage[version=3]{mhchem}

\usepackage{tabularx}

\author{Ruiyu Wang}
 \affiliation{Institute for Physical Science and Technology, University of Maryland, College Park, MD 20742, United States}
\author{Pratyush Tiwary}
 \email{ptiwary@umd.edu}
 \affiliation{Institute for Physical Science and Technology, University of Maryland, College Park, MD 20742, United States}
 \alsoaffiliation{Department of Chemistry and Biochemistry, University of Maryland, College Park, MD 20742, United States}
 \alsoaffiliation{Institute for Health Computing, University of Maryland, Rockville, MD 20847, United States}

\title [title]{Atomic scale insights into NaCl nucleation in nanoconfined environments}

\begin{document}

\begin{tocentry}

    \centering
    \includegraphics[width=0.75\textwidth]{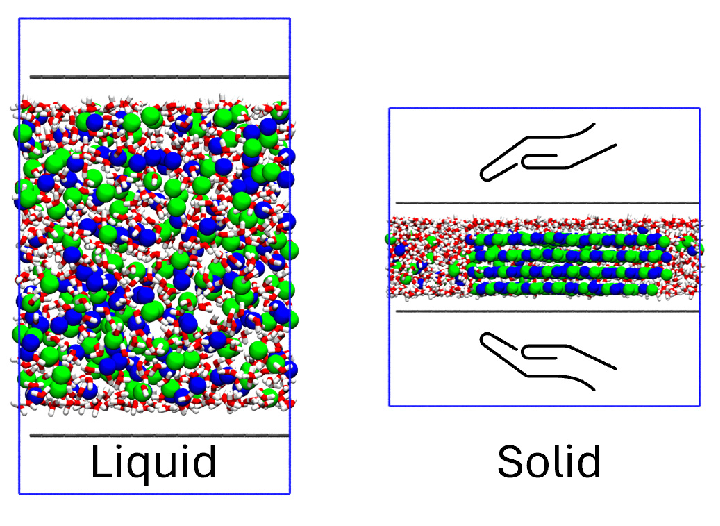}

\end{tocentry}

\begin{abstract}
In this work we examine the nucleation from NaCl aqueous solutions within nano-confined environments, employing enhanced sampling molecular dynamics simulations integrated with machine learning-derived reaction coordinates. Through our simulations, we successfully induce phase transitions between solid, liquid, and a hydrated phase, typically observed at lower temperatures in bulk environments. Interestingly,  while generally speaking nano-confinement serves to stabilize the solid phase and elevate melting points, there are subtle variations in the thermodynamics of competing phases with the precise extent of confinement. Our simulations explain these findings by underscoring the significant role of water, alongside ion aggregation and subtle, anisotropic dielectric behavior, in driving nucleation within nano-confined environments. This report thus provides a framework for sampling, analyzing and understanding nucleation processes under nano-confinement.
\end{abstract}

\section{Introduction}
Understanding the chemistry and physics of interfaces is essential to solving critical problems in climate, energy, and water \cite{bjorneholm_cr_2016,tang_cr_2020,gonella_nrc_2021,ruiz_nrc_2020,grob_cr_2022,calegari_jpcl_2023,le_jpcl_2021, striolo_arcbe_2016,wang_epj_2019,remsing_2015_pnas,banuelos_cr_2023,lee_pnas_2021}.
Of particular interest is planar nano-confined water, created by a water slab near two planar solid surfaces separated by several nanometers.
Such nano-confined environments have been extensively utilized in industry, material science, and biochemistry \cite{chakraborty_acr_2017}. They affect water oxidation \cite{bhullar_jmca_2021}, proton transfer \cite{munoz_prl_2017}, water wetting \cite{rego_arcmp_2022} and ion transport \cite{lynch_acsn_2021}. The confinement of water significantly modifies the physics and chemistry compared to bulk water \cite{munoz_cr_2021,ilgen_arpc_2023}. 

Consider for instance the dielectric constant, unlike an isotropic value of $\epsilon \approx 78$ in bulk water, $\epsilon$ perpendicular ($\epsilon_{\bot} \approx 2$) and parallel ($\epsilon_{\parallel} \approx 200$) to confining planar surfaces differ by at least one order of magnitude \cite{ruiz_pccp_2020,leung_nl_2023}. As a result, nano-confined water shows some universal trends that affect chemical reactions, such as inhibiting water self-dissociation, regardless of the type of surfaces and specifics of the surface-water interaction \cite{dipino_angew_2023}. Another interesting finding pertains to how water density distributions can show several pronounced layers between the nano-confining surfaces. 
The local density of interfacial water is strongly correlated with $\epsilon_{\parallel}$.
Interestingly, it is missing waters beyond solid surfaces, instead of the interaction between water and surfaces, that leads to the decrease of $\epsilon_{\bot}$. Thus the distance between the nano-confining surfaces is an important factor in tuning the dielectric behaviors of confined water 
and other liquids \cite{olivieri_jpcl_2021,motevaselian_acsn_2020}.

Although there is a long history of investigating the physics and chemistry of nano-confined water, 
the nature of phases and associated phase transitions, especially in solvated ionic systems under differing extents of nano-confinement is not as well studied.
Nucleation of new phases is difficult to study using experimental observations because of the small size of critical nuclei at the scale of a nano-meter.
One often one uses classical nucleation theory (CNT), a simplified model to estimate the nucleation rate.
It assumes a simple single-step mechanism wherein solute aggregation and nucleation occur simultaneously, but often fails to match other observations \cite{sosso_cr_2016,tsai_jcp_2019}. As an alternative to CNT, molecular dynamics (MD) simulations have been used to predict the relative stability of different phases. For instance, MD simulations have shown that the melting point of water under confinement is 100 K lower than that of bulk \cite{kapil_nature_2022}. However, the time scale of nucleation is much longer than the capability of typical MD, making simulations of the nucleation process impossible. One solution is use of enhanced sampling such as well-tempered metadynamics (WTMetaD) \cite{barducci_prl_2008,dama_prl_2014,bussi_nrp_2020}, which adds biased potential to help simulations escape from free energy minima to sample rare events. The quality of WTMetaD relies on the choice of the biasing variable, which should approximate the true \textit{a priori} unknown reaction coordinates (RCs) \cite{bussi_nrp_2020}. Recently the problem of designing biasing variables for enhanced sampling has seen progress with the use of machine learning (ML), that can approximate the RC from limited data. 
Here we use the state predictive information bottleneck (SPIB) approach \cite{wang_jcp_2021,mehdi_jctc_2022} that has been successfully applied to study the nucleation of urea \cite{zou_pnas_2023,zhao_jctc_2023}, iron \cite{zou_arxiv_2023} and NaCl \cite{wang_jpcb_2024}.

\begin{figure}[h!]
    \centering
    \includegraphics[width=0.425\textwidth]{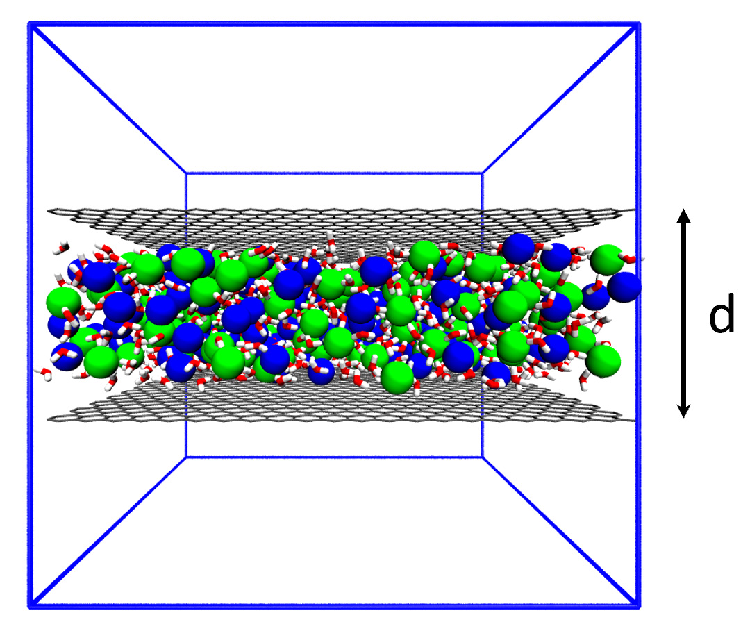}
    \caption{A snapshot of the simulation box. C, Na, Cl, O, and H atoms are indicated with colors grey, blue, green, red, and white, respectively. Blue lines show the boundary of the simulation box. Vacuum is reserved perpendicular to graphene sheets (grey color), separated by a tunable distance $d$.}
    \label{fig:boxtext}
\end{figure}

In this work, we study the nucleation of NaCl from aqueous solutions under nanoscale confinement using such ML-based enhanced sampling methods \cite{barducci_prl_2008, wang_jcp_2021, mehdi_arxiv_2022}.
Past investigations have made the intriguing, and not yet entirely explained observation that NaCl prefers to be solid instead of liquid \cite{zhao_nc_2021,zhao_jpcl_2022} showing hexagonal crystals inside graphene slits separated by a few nanometers \cite{wang_prl_2021}. Here we use ML-augmented MD to simulate this system for different extents of nano-confinement.
The confinement is tuned by changing the thickness $d$ of two graphene sheets to mimic sub-nano-confinement (Fig.~\ref{fig:boxtext}), maintaining the same NaCl concentration as shown in Table \ref{tab:sim_number} in section \textbf{Methods}. 
We apply ML in two different ways. First, using the SPIB approach we learn the nano-confinement dependent reaction coordindate as a neural network. Second, we quantify the importance of different molecular determinants or order parameters (OPs) in the nucleation process by using the Thermodynamically Explainable Representations of AI and other black-box Paradigms (TERP) approach \cite{mehdi_arxiv_2022}.  

We investigate the influence of confinement thickness on nucleation, aiming to quantify and better understand the nature of transition between confinement-like and bulk-like behaviors. In our WTMetaD simulations, we observed transitions among liquid, solid, and other uncommon phases. Varying the confinement thickness revealed that generally speaking confinement promotes crystallization, including the formation of a hydrated structure typical at low temperatures, akin to raising melting points. We demonstrate how this change likely stems from distinct dielectric behavior of water under confinement at the nanoscale. We also show that specific ion-water interactions, rather than just ion structures, are crucial to understanding the process. 
This stems in the observation that in nano-confined aqueous solutions, the nucleation of NaCl requires removing solvent water at the surface of solid nucleus.
Our work introduces a generic protocol for simulating and analyzing nucleation, providing insights into chemical and physical processes at the nanoscale.

\section*{Results and Discussions}

\subsection*{NaCl Phases under Nanoconfinement}

\begin{figure*}
    \centering
    \includegraphics[width=0.93\textwidth]{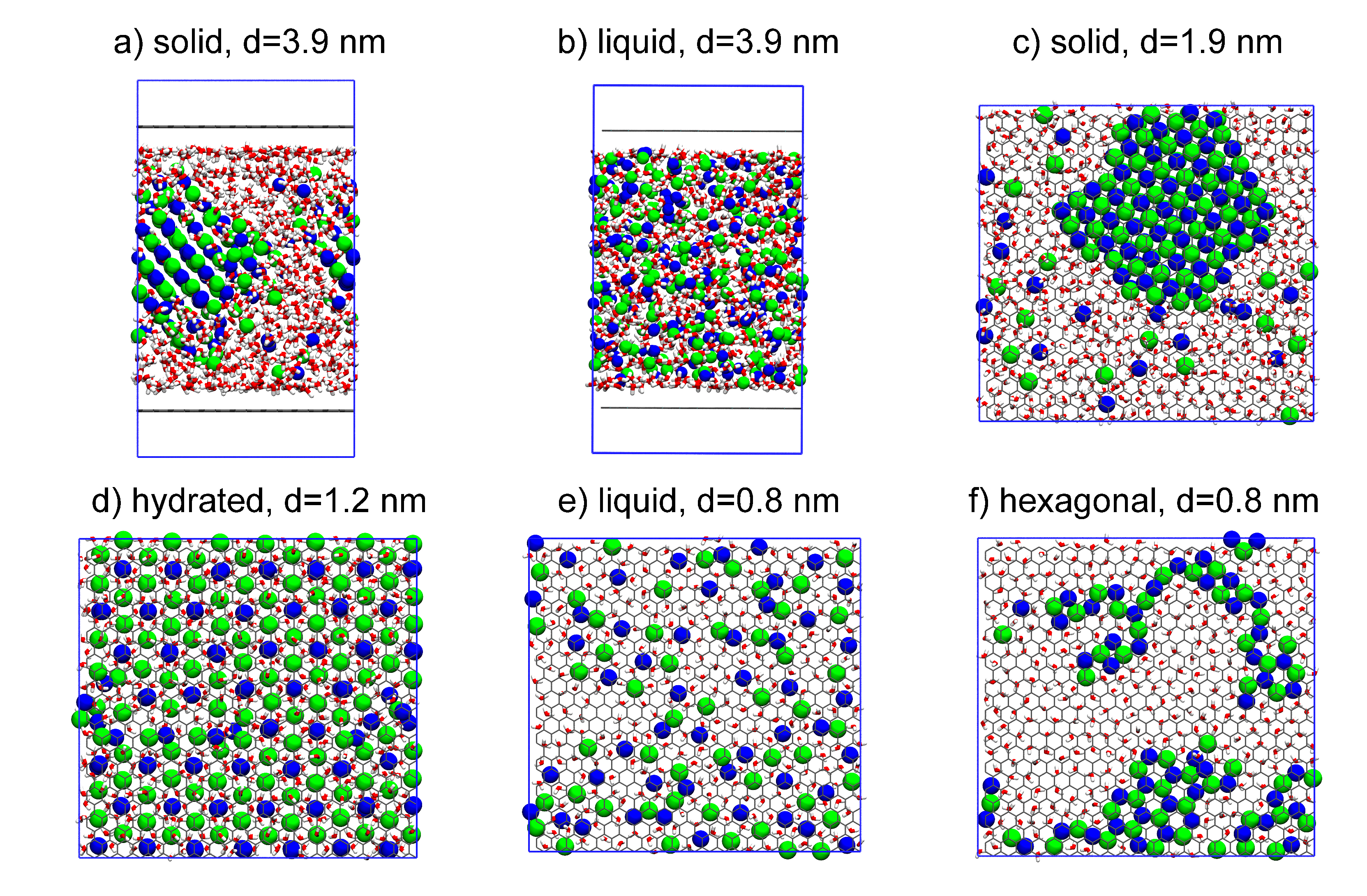}
    \caption{ Snapshots of phase structures of NaCl aqueous solutions observed. $d$ is the distance between two graphene sheets as discussed above. Snapshots (c) to (f) are snapshots from the top of graphene sheets. Not all ions on (c) and (d) are on the same plane.}
    \label{fig:textcrystal}
\end{figure*}

We first report results for simulations of NaCl in bulk water under room temperature, which corresponds to $d = 3.9$ nm. See Supplementary Materials (SM) for further details of the simulation set up). Here, in accordance with previous simulations and experiments, we find only one crystalline structure for NaCl: the face-centered cubic (FCC) solid with a coordination number of 6 for counterions forming an octahedron (Fig.~\ref{fig:textcrystal}a). 
The solid and liquid structures (Fig.~\ref{fig:textcrystal}a and b) are not different from simulations without any graphene sheets \cite{zimmermann_jacs_2015,jiang_jcp_2018,jiang_jcp_2019,giberti_jctc_2013}.
Reducing $d$ does not affect the growth of the solid parallel to surfaces, but it becomes prohibited in perpendicular direction, limiting the thickness of the solid to only several ($d=1.9$ and $1.2$ nm) layers. Under stronger confinement, we find that additional crystalline structures start to emerge. These include the hydrated structure, which is similar to the hydrohalite NaCl.2H2O (Fig.~\ref{fig:textcrystal}d) at $d=1.2$ nm, and hexagonal (Fig.~\ref{fig:textcrystal}f) at $d=0.8$ nm.
In the hydrated NaCl structure, Na$^+$ and Cl$^-$ ions are not in direct contact with other ions, but are connected through bridge waters. In experiments, this structure has been reported to usually exist below 0.1 \textdegree C, but has also been observed under confinement at room temperature, indicating that nano-confinement promotes crystallinity and increases melting point\cite{bode_cgd_2015}. 
However, further increasing the confinement leads to the disappearance of hydrated structures. The hexagonal (Fig.~\ref{fig:textcrystal}f) crystal structure \pta{that we observe in the vicinity of the graphene} has also been observed in experiments \cite{wang_prl_2021,tikhomirova_jpcl_2020}, as an intermediate before the formation of solid crystals.  The hexagonal crystal is only observed when $d=0.8$ nm and it does not align with the graphene structure \cite{wang_prl_2021}.

\subsection*{\rya{Nanoconfined Environments Stabilize Solid Phases}}

\begin{figure*}
    \centering
    \includegraphics[width=0.95\textwidth]{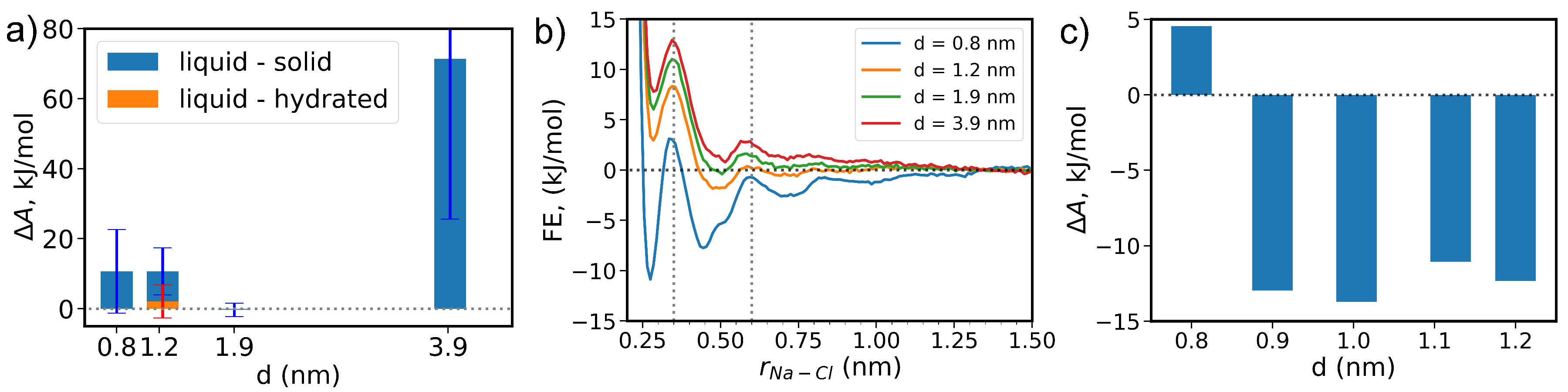}
    \caption{  Free energy plots of phase transition and ion pairing. (a) The relative free energy difference of liquid-solid (blue) and liquid-hydrated (yellow bar and red error bar) phases of NaCl solution under nano-confinement. There is only one yellow bar because the hydrated phase is only observed at $d=1.2$ nm.
    (b) Free energy surfaces for a single ion-pair dissociation. The vertical dashed lines at 0.35 and 0.6 nm mark middle of the basin of the contact pair (CP) and solvent-separated ion-pair (SSIP) respectively. (c) The relative free energy of the CP and SSIP states for a single ion-pair for $d$ between 0.8 to 1.2 nm. }
    \label{fig:festate}
\end{figure*}

We report the relative free energy differences between liquid and solid phases calculated for all thicknesses (Fig.~\ref{fig:festate}a), obtained by reweighting the metadynamics simulations \cite{tiwary_jpcb_2015}.
The positive $\Delta A$ for all thicknesses show that the liquid phase with dissolved NaCl is the most stable phase.
Simulations with the largest thickness, $d=3.9$ nm show the highest free energy around 70 $kJ/mol$, showing that the liquid phase is most stable as bulk conditions are approached, 
consistent with our previous observations \cite{wang_jpcb_2024}. When decreasing $d$, while liquid is still the most stable phase, the relative free energy benefit is now less than 10 $kJ/mol$. The lower free energies demonstrate that nano-confinement promotes the formation of the solid. Interestingly, the behavior with $d$ is non-monotonic and non-trivial, as the most stable solid appears at $d=1.9$ nm, where all phases are approximately equally stable.
Further reducing $d$ decreases the relative stability of the solid phase. We attribute the observation that intermediate $d$ leads to the most stable solid to \pta{ion} depletion during nucleation. Simulations with smaller $d$ contain \pta{fewer} molecules. Such stronger finite size effect destabilizes the solid state due to depletion of ions \cite{li_jpcl_2023}. 
The hydrated phase is \pta{thus} more favored than the solid at $d=1.2$ nm, but is not seen for other nano-confinements because in hydrated NaCl, ions do not lie on the same plane but at $d=0.8$ nm, the space is insufficient for ions to form two-layer configurations.

To gain further insight into the relative free energy between different phases, we study the dissociation of a single ion-pair under differing amounts of nano-confinement (Fig.~\ref{fig:festate}b). For thickness $d=3.9$ nm, ions prefer to be dissociated, consistent with the fact that liquid is more stable than solid. 

Reducing $d$ makes solvent-separated ion-pair (SSIP) and contact pair (CP) more favored than that in bulk for all thicknesses. For $d \leq 1.9$ nm, SSIP is preferable over dissociated ions. The change in the free energy surface (FES) for $d=0.8$ nm is much more significant than that of other thicknesses (Fig.~\ref{fig:festate}b), as the CP is more stable than SSIP \pta{only for the case of $d=0.8$ nm}.

For thickness $d=1.2$ nm, ion pairing and nucleation show exactly the same trend: SSIP is more favored than CP and the hydrated phase is only observed for this thickness. We expect this is because a periodic ordered structure may stabilize the hydrated phase (Fig.~\ref{fig:textcrystal}d). However, ion pairing and nucleation are not always the same. For instance, for thickness $d=1.9$ nm, unpaired and SSIP have significantly lower free energies than the CP. On this basis, one would expect that the solid should not be preferred, while \pta{our nucleation simulations show that} for $d=1.9$ nm, the solid phase is as stable as the liquid phase (Fig.~\ref{fig:festate}a). Lastly, only in simulations for thickness $d=0.8$ nm, CP is more favored than SSIP. The free energies of both CP and SSIP are lower than unpaired ($r > 1.5$ nm), which is consistent with previous observation that small solid nuclei can be observed even in unbiased simulations \cite{zhao_nc_2021}.

We can explain the above contradictions on the following basis besides the depletion and finite size effects. There are no such effects in single ion-pairing. The high initial concentration of NaCl solution further decreases the dielectric constant of the solution to stabilize solid crystals. The $\epsilon$ \pta{for} NaCl aqueous solution in the absence of surfaces at the concentration of this work is only half of that of bulk pure water \cite{zhang_prl_2023}. Reduced $\epsilon$ elevates the Coulomb forces between ions, which is believed to be the driving \pta{force for the formation of} the solid phase.

We also notice a sharp change in the FES for $d=0.8$ and $1.2$ nm \pta{(Fig.~\ref{fig:festate}b)}. \pta{To better explore this, we} calculate the relative stability between CP and SSIP states for additional thicknesses $d$ \pta{in between $d=0.8$ and $1.2$ nm}. We observe a discrete change in the state stability. CP is favored only at $d=0.8$ and \pta{reducing the nanoconfinement by} even 0.1 nm reverses the trend. Since the dielectric constant varies continuously with the thickness $d$ \cite{ruiz_pccp_2020,munoz_cr_2021},  the discontinuity in states stability may be affected by water layering. There are 3 and 2 water layers for $d=1.2$ and $0.8$ nm, respectively.

\subsection*{\rya{Ion Dehydration and Ion-water Electric Forces Drive Nucleation}}

Though with the use of SPIB we can distinguish different stable states \cite{wang_jcp_2021}, \pta{we now perform further analyses to understand the reaction coordinate learnt by SPIB. Specifically, to better understand what drives nucleation, we focus only in the vicinity of the liquid-solid transition regions. We perform such a local analysis because SPIB uses global} data as the input \cite{wang_jpcb_2022}.
For this additional analysis here we use the Thermodynamically Explainable Representations of AI and other black-box Paradigms (TERP) method that focuses locally near the liquid-solid transition states to evaluate the importance of OPs that drive the nucleation \cite{mehdi_arxiv_2022}.

Among selected OPs, one category with high TERP scores is the number of ions with high coordination number of counterion, $N_{x+}$.
$x$ is selected as 4 for square lattice in $d=0.8$ nm and 5 for FCC lattice in other thicknesses.
The OP is an approximation of number of ions in the core of solid, but it is continuous and can be biased in WTMetaD simulations. For $d=3.9$ nm, $N_{5+}$, together with Steinhardt bond OP $\overline{q_4}$, have high TERP score, consistent with previous result in NaCl solution without graphene sheets \cite{wang_jpcb_2024}. 

\begin{figure*}[h!]
    \centering
    \includegraphics[width=0.95\textwidth]{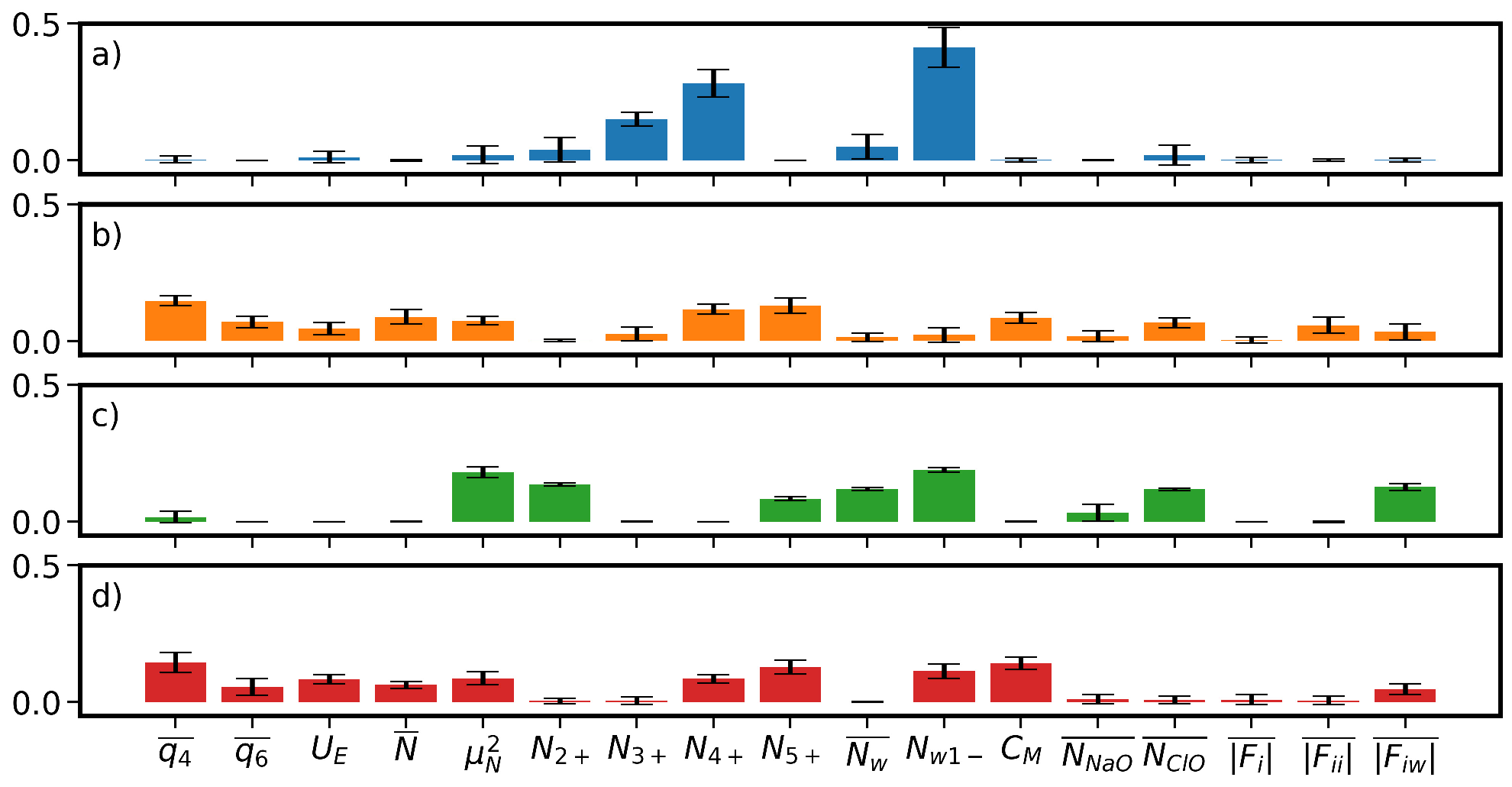}
    \caption{The average TERP score (Y axis) of selected points at the transition state of liquid and solid phase transition for all OPs (X axis) at thicknesses (a) $d=0.8$ nm, (b) $d=1.2$ nm, (c) $d=1.9$ nm and (d) $d=3.9$ nm. Error bars represent one standard deviation. }
    \label{fig:terp}
\end{figure*}

\pta{With TERP we are able to quantify the role of water in nucleation of NaCl from aqueous solution, which} is often overlooked or not quantitatively characterized \cite{finney_wires_2023,sosso_cr_2016, bolhuis_arpc_2002}.  Here, among water-involved OPs, the number of ions with one or less coordination water, $N_{W1-}$, has a significantly high TERP score except for $d=1.2$ nm. It is the highest for $d=0.8$ and $1.9$ nm (Fig.~\ref{fig:terp}), indicating that it is the most important OP for nucleation under nano-confinement.
The OP is highly correlated with $N_{4+}$ or $N_{5+}$ (Fig.~S9) because coordinating with a counterion requires removing a water molecule. However, $N_{W1-}$ explicitly includes water. \ry{The growth of $N_{W1-}$ relies on the removal of water from the nucleus but randomly removing coordinated waters does not drive the nucleation (corresponding to the OP average coordinated water, $\overline{N_W}$), 
whereas the growth of $N_{5+}$ is just to move ions to nucleus. }
We further investigate the role of hydration of \ce{Na+} and \ce{Cl-} because both MD simulations \cite{mukhopahyay_jpcb_2012, zhang_pccp_2020,thiemann_acsnano_2022,ding_pnas_2014} and experiments \cite{gonella_nrc_2021, montenegro_nat_2021} show that water responds asymmetrically to charged planar solid surfaces. TERP results show that the hydration \ce{Cl-} (represented by the average number of oxygen contact with \ce{Cl-}, $\overline{N_{ClO}}$) plays a greater role in driving the nucleation 
\ry{because water bonds tighter with \ce{Cl-}, making it difficult to remove water from \ce{Cl-}, in addition to a larger size of its hydration shell.}

As previously mentioned, the Coulomb forces between ions and between ions and water molecules are believed to be the primary drivers of nucleation and dissolution \cite{zhao_jpcl_2022}. Our analysis reveals that TERP predicts a non-zero score for ion-water forces, $|F_{iW}|$ (Fig.~\ref{fig:terp}), which is consistent with the high TERP score of $N_{W1-}$. This is thus evidence that alterations in the dielectric constant within a nano-confined environment influence the Coulomb forces between ions and water, consequently impacting the nucleation of NaCl. However, TERP scores associated with electric force OPs are not notably high, indicating that these OPs require further refinement to accurately depict the nucleation processes.

We attribute the observation that nano-confined environment stablizes and increases the melting point of the solid state to the entropy \pta{effects due to} geometry. The nano-confined environment restricts ion movement and reduces the entropy contribution to the free energy of phase transition. As a result, a higher temperature from the $T\Delta S$ term is needed. Effects on enthalpy by nano-confinement are supposed to be relatively smaller than entropy.

\section*{Conclusion}

In this work, we have investigated the nucleation of NaCl aqueous solution under nano-confinement using ML-based enhanced sampling molecular dynamics simulations. 
The nano-confinement is tuned by changing the distance between two graphene sheets containing NaCl solutions.
Machine learning is applied to extract reaction coordinates that successfully drive the phase transition, including to common solid and liquid phases and uncommon ones such as hydrated phase that is supposed to exist at low temperatures in the bulk. Free energy analysis shows that nano-confinement stabilizes the solid phase, equivalent to raising the melting point. 
We \pta{also explored} the connection between nucleation and single ion pairing. In bulk-like solutions, the liquid phase and unpaired ions are preferred, and strong confinements make solid and the contact ion pair favored. Reducing the thickness of water slabs leads to discrete change in ion pairing, indicating the role of water layering in ion pairing and nucleation. 
We evaluate the driving force for nucleation by calculating the importance of order parameters. 
Under nano-confined environment, the removal of interfacial water of the nucleus of the solid phase, especially water that contact with \ce{Cl-}, as well as the electric force between ion and water, are considered to drive the nucleation.
We expect that the results of this work could provide better insight into the investigation of processes under nano-confined environment or at solid/liquid interfaces. The role of solvents, either their collective behavior or local structures, is an important factor to be considered. 
The results \pta{we have provided in this work could also} provide insights for \pta{further applications, including} the design of energy materials. Consider the instance of water dissociation, an important step of electrocatalysis \cite{dipino_angew_2023,wang_jpcc_2019,wang_wcms_2021,xu_jacs_2024}, we assume a general trend that planar nano-confined environment suppresses charge separation due to the change of dielectric behavior and hydration structure of ions from bulk water. 
The ability to tune specific interactions between solutions and surfaces \pta{driven by such atomic scale thermodynamic and mechanistic insights could have a pronounced role in the} the design of materials.

\section{Methods}
The simulation settings were based on our previous work \cite{wang_jpcb_2024}. 
All MD simulations were carried out using GROMACS 2022.3 \cite{gromacs_1995,gromacs_2015}. All simulations were performed using the constant number, volume, and temperature (NVT) ensemble using a time step of 2 fs. The temperature was maintained at 300 K using \rya{canonical sampling through velocity rescaling} with a relaxation time of 0.1 ps \cite{nvt}.
\ry{The ratio of NaCl and water molecules is about 1:5.287, corresponding to a bulk concentration of 8.86 mol/L,}
which is around 1.5 times of the saturated concentration (denoted as $1.5c_s$). 
The Joung-Cheatham force field is for NaCl\cite{zimmermann_jacs_2015,jc_ff}. The graphene surfaces were built using CHARMM-GUI Nanomaterial Modeler and the INTERFACE force field is applied \cite{charmmgui,charmm_gui_carbon_ff}, and kept fixed during simulations.  The SPC/E model was applied to water \cite{berendsen_jpc_1987}.
Water OH bonds were fixed using the LINCS algorithm \cite{lincs}.  The cutoff of short-range interactions was 1 nm and long-range electrostatic interactions were calculated using particle-mesh Ewald summations \cite{ewald}.  Periodic boundary conditions were applied in all XYZ directions.  

Enhanced sampling was carried out using metadynamics (MetaD) with PLUMED package version 2.8.1 \cite{plumed,bonomi_nm_2019}. Bias potentials with an initial height of 5 kJ/mol were added to simulations every 2 ps. The transition tempered MetaD was applied in simulations the using SPIB1 reaction coordinates (RC) for thickness $d=1.9$ and $3.9$ nm \cite{dama_jctc_2014}. The liquid and solid states are used as target states and the bias factor was set to 75. Well tempered MetaD (WTMetaD) is applied in other simulations and the bias factor was set to 100.

\subsection*{Simulation box}

An example snapshot of simulation in this work is shown in Fig.~\ref{fig:boxtext}. Other parameters are listed in Table~\ref{tab:sim_number}. Two parallel planar graphene sheets extend to X and Y directions with a fixed thickness (d) and the NaCl solutions are between the two sheets. For the simulation for $d=3.9$ nm, we reduced the simulation box to about $3 \times 3 \times 5$ ($nm^3$), since a larger simulation box contains too many ions and reduces the speed in enhanced sampling simulations, 

\begin{table*}

\caption{
Other parameters for the simulations. The units for thickness and box size are nm and $nm^3$ respectively. NaCl and Water represent the number of NaCl and water molecules.}
\label{tab:sim_number} %
\begin{tabularx}{0.95\textwidth}{p{4cm}p{6cm}p{3cm}p{3cm}}
\hline
  Thickness (d)   &  Box size &     NaCl &  Water \\
\hline
0.80 (d=0.8) &   5.4274 $\times$ 5.1276 $\times$ 5.2000  & 50 & 264 \\
1.23 (d=1.2) &  5.4274 $\times$ 5.1276 $\times$ 5.2000 &  110 & 580 \\
1.90 (d=1.9) &  5.4274 $\times$ 5.1276 $\times$ 5.2000 &  206 & 1088 \\
3.92 (d=3.9) &  2.9604 $\times$ 2.9911 $\times$ 5.2000 &  156 & 828\\
\hline

\end{tabularx}%
\end{table*}

\subsection*{Order parameters}
17 order parameters (OPs) are used to describe the structure of simulations. OPs that are biased in WTmetaD simulations must be continuous, including the average 4th ($\overline{q_4}$) and 6th ($\overline{q_6}$) Steinhardt Bond OPs, the average ($\overline{N}$), second moment ($\mu_N^2$) of all coordination number (CN) of Na-Cl, and the number of ions with CN more than 2, 3, 4, and 5 ($N_{2+}$, $N_{3+}$, $N_{4+}$, and $N_{5+}$) have been described in previous papers \cite{wang_jpcb_2024}. Here, we also introduce the following new OPs. 

\subsubsection*{Electric potential energy}
Previous results demonstrate that the thermodynamical descriptor enthalpy is effective in driving the nucleation of NaCl and other materials from the melt.\cite{piaggi_prl_2017,wang_jpcb_2024} Due to its implementation, it is not effective to drive the nucleation of NaCl from aqueous solution. Since the enthalpy is composed of Lennard-Jones and charge-charge interactions and the latter determines the structure of NaCl in aqueous solutions, 
we assume that biasing the electric potential energy of ions ($U_E$) could lead to the phase transition of NaCl. To implement the OP,  we use the DHENERGY keyword in PLUMED with EPSILON=1 and I=0.0001. 

\subsubsection*{Ion-water coordination numbers}
Although the ion-water interactions is often neglected in previous simulations of nucleation, it has been proved that the coordination number of ion-water is the reaction coordinate (RC) of ion pairing\cite{bolhuis_arpc_2002} and ligand binding\cite{beyerle_jpcb_2024}. The coordination number of an ion-water is defined by:
\begin{equation}
  s_w(i)=\sum_j \frac{1-(r_{ij}/r_0)^6}{1-(r_{ij}/r_0)^{12}}
\end{equation}
where $i$ and $j$ represent each ion and water oxygen atom. The distance $r_0$ for Na-O and Cl-O are 0.321 and 0.402 nm. To reduce the collection numbers to some scalars, we calculate the average ion-water coordination number of Na, Cl, and all ions ($\overline{N_{NaO}}$, $\overline{N_{ClO}}$ and $\overline{N_W}=0.5 \times \overline{N_{NaO}}+0.5 \times \overline{N_{ClO}}$). Similar to $N_{5+}$ for ion-ion CN, we believe that the total CN of an ion, with both counterions and water, is supposed to be a constant. Ions with more contact counterions have less coordinated water. As a result, the number of ions with coordinated water less than 1 ($N_{W1-}$), which is an approximation of ions belonging to the solid phase, is similar to $N_{5+}$ and the contribution of water is explicitly included.

\subsubsection*{Largest ion cluster}
For the NaCl concentration used in this work, the nucleation follows the one-step mechanism described in the classical nucleation theory (CNT)  in bulk water \cite{finney_wires_2023,finney_fd_2022}. In other words, the size of the largest ion cluster ($C_M$, \rya{represented by the number of ions in the cluster}) is another approximation to the size of the crystal and there are no disordered clusters and it is a widely used OP in MD simulations \cite{karmakar_jctc_2019}. 

\subsubsection*{Electric forces}
Previous papers show that the thickness of a water slab under confinement significantly affects the dielectric constant of the solution. The role of water, as the dielectric medium, is to screen the electric interaction of ions.
However, calculating the dielectric constant of a water slab in the presence of moving ions is not convenient \cite{bonthuis_prl_2011,zhang_prl_2023}.  At room temperature, the dissolution of NaCl in water is attributed to the solvation that undermines coulomb interaction between ions. Reducing the dielectric constant is equivalent to increasing their electric force and promoting the formation of solid phases. 
As a result, we believe that the dielectric behavior of the water slab can be described approximately using the electric forces of ions.
For electric forces OPs, we calculate the average of the norm of electric forces of all ions in each configuration for such 3 components: ion-(ion+water) ($|F_i|$), ion-ion ($|F_{ii}|$) and ion-water ($|F_{iw}|$).

\subsection*{Data Availability}
\begin{sloppypar}
Files to reproduce simulations in this work are available on GitHub (simulation settings) and PLUMED NEST (WTMetaD) at https://github.com/ruiyuwangwork/xxxxxx and https://www.plumed-nest.org/eggs/24/xxx/.

The data that support the findings of this study are available from the corresponding author upon reasonable request.
\end{sloppypar}

\begin{acknowledgement}

This research was entirely supported by the US Department of Energy, Office of Science, Basic Energy Sciences, CPIMS Program, under Award DE-SC0021009. We thank UMD HPC’s Zaratan and NSF ACCESS (project CHE180027P) for computational resources. R.W. thanks Tiwary group members for fruitful discussions and Shams Mehdi, Dedi Wang, and Dr. Eric Beyerle for helping prepare the manuscript.

\end{acknowledgement}

\begin{suppinfo}

Simulation details, additional analysis can be found in the supporting information.

\end{suppinfo}

\bibliography{achemso-demo}

\end{document}